\documentstyle[nato,numreferences,epsfig]{crckapb}

\newcommand{\dual}{\mbox{}^{\ast}}
\newcommand{\dd}{\mbox{\rm d}}
\newcommand{\dD}{{\cal D}}
\newcommand{\Z}{{Z \!\!\! Z}}
\newcommand{\beq}{\begin{equation}}
\newcommand{\eeq}{\end{equation}}
\newcommand{\beqn}{\begin{eqnarray}}
\newcommand{\eeqn}{\end{eqnarray}}
\newcommand{\eq}[1]{(\ref{#1})}
\newcommand{\cD}{{\cal D}}

\newcommand{\cZ}{{\cal Z}}

\newcommand{\preprint}{\vspace{-13.3cm}
 \rightline{\small ITEP-LAT/2002-05}
 \rightline{\small KANAZAWA-02-08}
 \vspace{1mm}
 \rightline{\small 30 April, 2002}
 \vspace{11.6cm}}
\newcommand{\slovakia}{\vspace{-6mm}
 \leftline{\footnotesize{* Presented by the third author
 at the NATO Advanced Research Workshop}}
 \vspace{-1mm} \leftline{\footnotesize{
 "Confinement, Topology, and other Non-Perturbative Aspects of
 QCD",}}
 \vspace{-1mm} \leftline{\footnotesize{
 Star\'a Lesn\'a, High Tatra Mountains (Slovakia),
 January 21-27, 2002}} \vspace{3mm}}

\begin{opening}
\title{CONFINEMENT - DECONFINEMENT ORDER PARAMETERS}
\author{V. A. Belavin${}^{\lowercase{a}}$}
\author{M. N. Chernodub${}^{\lowercase{a,b}}$}
\author{M. I. Polikarpov${}^{\lowercase{a}}$}
\institute{${}^{a}$ Institute of Theoretical and Experimental
Physics\\ B. Cheremushkinskaya, 25, Moscow, 117259, Russia\\
${}^{b}$ Institute of Theoretical Physics, University of Kanazawa,
\\ Kanazawa 920-1192, Japan}

\end{opening}

\runningtitle{Confinement--deconfinement order parameters}

\begin{document}

\begin{abstract}
We study numerically the monopole creation operator proposed
recently by Fr\"ohlich and Marchetti. The operator is defined with
the help of a three dimensional model which generates random
Mandelstam strings. These strings imitate the Coulombic magnetic
field around the monopole. We show that if the Mandelstam strings
are condensed the creation operator discriminates between the
phases with condensed and non--condensed monopoles in the Abelian
Higgs model with the compact gauge field.
\end{abstract}

\slovakia\preprint

\section{Introduction}

The order parameters are very important for the investigation of
the phase transitions. In case of the temperature phase transition
in lattice QCD with dynamical quarks there is no good definition
of the order parameter up to now. The traditional order parameters
like Polyakov line and string tension which work well in the
quenched case are not valid in case of the full QCD. Even at zero
temperature the string between quark and anti-quark can be broken
by sea quarks.

We discuss below a quantity which may serve as the order parameter
for full QCD if the monopole (or, "dual superconductor")
confinement mechanism~\cite{DualSuperconductor} is valid. In this
picture the monopoles -- defined with the help of an Abelian
projection~\cite{AbelianProjections} -- are supposed to be
condensed in the confinement phase. The monopole condensate causes
a dual analogue of the Abrikosov vortex to be formed between
quarks and anti-quarks. As a result the quarks and anti--quarks
are confined into the colorless states. In the deconfinement phase
the monopoles are not condensed and quarks are not confined. Thus
the natural confinement-deconfinement order parameter is the value
of the monopole condensate, which should be nonzero in the
confinement phase and zero in the deconfinement phase.

There are two (formal) difficulties in the definition of the
monopole condensate. At first, the monopole condensate is defined
as an expectation value of the monopole field. However, the
monopoles are the topological defects in the compact Abelian gauge
field and, as a result, the immediate output of the lattice
simulations is an information about the monopole trajectories.
Then one can apply a known procedure which
allows us to rewrite the path integral over monopole trajectories as
an integral over the monopole fields. The expectation value of the
latter defines the monopole condensate.

The next difficulty is the following. The expectation value of the
({\em e.g.}, scalar) field $\phi$ should always be zero regardless
whether this field is condensed or not. The reason is very simple:
the path integral includes the integration over all possible
gauges while the charged field is gauge dependent. These two
problems were solved in paper \cite{FrMa87}, where the gauge
invariant monopole creation operator for compact QED (cQED) was
explicitly constructed. The numerical calculations in lattice cQED
\cite{PoWi} and in the Maximal Abelian projection of
lattice $SU(2)$ gluodynamics, Ref.~\cite{ChPoVe}, show that this
operator provides the order parameter for confinement --
deconfinement phase transition. Later the other forms of the
monopole creation operators were constructed and investigated
numerically \cite{DiGi} in lattice gluodynamics.

But recently it appears the claim \cite{FrMa99} that the monopole
creation operator suggested in \cite{FrMa87} depends on the shape
of the Dirac string if dynamical electrically charged fields are
present in the considered theory. Just this situation appears in
the Abelian projection of QCD: the off-diagonal gluons become
electrically charged dynamical fields while the diagonal gluons
become compact Abelian gauge fields which contain monopoles. The
authors of \cite{FrMa99} suggested a ``new'' monopole creation
operator, which does not depend on the shape of the Dirac string
even in the presence of the dynamical electric charges.

Below we study this new creation operator in the compact Abelian
Higgs model, having in mind the future application of this operator for
the full QCD.

In the next Section we give the explicit construction of the ``new''
and the ``old'' monopole creation operator. In Section 3 we present the
results of the numerical calculations in the compact Abelian Higgs
model.

\section{``Old'' and ``new'' monopole creation operators.}

The gauge invariant creation operator $\Phi$ was suggested by
Dirac \cite{Dirac}:
\beqn
\label{Dirac}
\Phi = \phi (x) \exp\left\{i \int E_k(\vec{x}-\vec{y}) A_k(\vec{y}) \,
d^3 y\right\} \,\, ,
\eeqn
here $\phi(x)$ and $A_k (x)$ are the electrically charged field and
gauge potential,
which transform under the gauge transformations as
\beqn
\label{GT}
\phi (x) \to \phi (x) e^{i\alpha(x)}\,,\qquad
A_k(x) \to A_k(x) + \partial_k \alpha(x)\,.
\eeqn

The Coulomb field, $E_k(x)$, satisfies the equation:
\beqn
\partial_k E_k = \delta^{(3)} (x)\, .
\eeqn
It is easy to see that the operator $\Phi$, eq.~\eq{Dirac}, is invariant under
the gauge transformations \eq{GT}.

Now we describe the Fr\"ohlich--Marchetti construction
\cite{FrMa87} of the monopole creation operator in cQED. At first
step the partition function of cQED is transformed to a dual
representation. For the general form of the cQED action it can be
shown \cite{ChPoVe} that the dual theory is an Abelian Higgs model
(AHM) in the limit when the Higgs boson mass and the gauge boson
mass are infinite. In this theory the Higgs field, $\phi_x$,
corresponds to the monopole in the original cQED. The gauge field
$\dual B$ is dual to the original gauge field $\theta$. Thus the gauge
invariant creation operator \eq{Dirac} for the AHM model,
corresponds to the monopole creation operator in the original
cQED. The explicit expression for this operator on the lattice is
({$cf$. eq.\eq{GT}}):
\beq
\Phi^{\mathrm{mon}}_x = \phi_x \, e^{i (\dual B, \dual H_x)}\,,
\label{Phi}
\eeq
where $\dual H_x$ is the Coulomb field of
the monopole, $\delta \dual H_x = \dual \delta_x$, and
$\dual \delta_x$ is the discrete $\delta$--function defined on the
dual lattice. Here and below we will use the differential
form notations on the lattice: $(a,b) = \sum_c a_c b_c$  is the
scalar product of the forms $a$ and $b$ defined on the $c$--sells;
$(a,a) \equiv {||a||}^2$ is the norm of the form $a$; $d$
is the forward derivative (an analog of the gradient);
$\delta$ is the backward derivative (an analog of the divergence)
and $\delta$--operation transfers a form to the dual lattice.
For a description of the language of the differential forms on
the lattice see, $e.g.$, review~\cite{Review}.

Performing the inverse duality transformation for the expectation
value of the creation operator \eq{Phi} we get
the expectation value, $\langle\Phi^{mon}\rangle$, of this operator
in cQED. The explicit expression in the lattice notations is:
\beqn
\langle \Phi^{\mathrm{mon}} \rangle \! & = & \! {1 \over \cZ}
\int\nolimits_{- \pi}^{\pi} \!\!\! \dD \theta \, \exp\{ -S(\dd \theta + W)\}\,,
\nonumber\\
\cZ \! & = & \! \int\nolimits_{- \pi}^{ \pi} \!\!\! \cD \theta \, \exp \{-
S(d\theta) \}\,, \label{original}
\eeqn
here $\dd\theta$ is the plaquette angle, the lattice action is a
periodic function: $S(\dd\theta+2\pi n)=S(\dd\theta)$, $n \in \Z$; $W=2
\pi \delta \Delta^{-1}\dual (H_x - \omega_x)$ and $\dual \omega_x$ is the
Dirac string which starts at the monopole: $\delta \dual \omega_x =
\dual \delta_x$. The Dirac string $\dual \omega_x$ is defined on the dual lattice.
 The numerical investigation of this creation operator
in cQED shows \cite{PoWi} that it can be used as the
confinement--deconfinement order parameter.

The operator \eq{Phi} is well defined for the theories without
dynamical matter fields. However, if an electrically charged
matter is added, then the creation operator \eq{Phi} depends on
the position of the Dirac string. To see this fact
let us consider the compact Abelian Higgs model with
the Villain form of the action:
\beqn
\cZ_{AHM} = \int^\pi_{-\pi} \!\!\!\dD \theta \int^\pi_{-\pi} \!\!\!\dD  \varphi
\sum_{n \in \Z(c_2)} \sum_{l \in \Z(c_1)} \,
e^{- \beta ||{\mathrm d} \theta + 2\pi n||^2
- \gamma || {\mathrm d} \varphi + q \theta + 2\pi l||^2}\,.
\label{ZAHM}
\eeqn
Here $\theta$ is the compact Abelian gauge field and $\varphi$ is
the phase of the dynamical Higgs field. The integer $q$ is the charge of
the Higgs field. For the sake of simplicity we consider the London
limit (the Higgs mass is infinitely large while the Higgs condensate
is constant).

Let us perform the Berezinsky-Kosterlitz-Thouless (BKT)
transformation~\cite{BKT} with respect to the compact gauge field
$\theta$:
\beqn
\dd \theta + 2 \pi n = \dd A + 2 \pi \delta \Delta^{-1} j\,, \quad
\mbox{with} \quad
A = \theta + 2 \pi \delta \Delta^{-1} m[j] + 2 \pi k\,.
\label{BKTtheta}
\eeqn
Here $A$ is the non--compact gauge field, $\dual m[j]$ is a surface on the
dual lattice spanned on the monopole current $\dual j$ ($\delta
\dual m[j] = \dual j$), $\Delta$ is the lattice Laplacian and $k$
is the integer--valued vector form\footnote{A detailed description
of the duality and BKT transformations in terms of the
differential forms on the lattice can be found, $e.g.$, in
Ref.~\cite{Review}.}. We substitute eqs.\eq{BKTtheta} in
eq.\eq{ZAHM} and make the shift of the integer variable, $l \to l +
q k$.

Next we perform the BKT transformation with respect to the
compact scalar field $\varphi$:
\beqn
\dd \varphi + 2 \pi l = \dd \vartheta + 2 \pi \delta \Delta^{-1} \sigma\,,
\quad \mbox{with} \quad
\vartheta = \varphi + 2 \pi \delta \Delta^{-1} s[\sigma] + 2 \pi p\,.
\label{BKTphi}
\eeqn
Here $\vartheta$ is the non--compact scalar field, $\dual s[\sigma]$ is
a $3D$ hyper--surface on the dual lattice spanned on the closed surface
$\dual \sigma$ ($\delta \dual s[\sigma] = \dual \sigma$) and $p$
is the integer--valued scalar form.

Substituting eqs.~(\ref{BKTtheta},\ref{BKTphi}) into the partition
function~\eq{ZAHM} and integrating the fields $A$ and $\varphi$
we get the representation of the compact AHM in terms of the monopoles
and strings ("the BKT--representation"):
\beqn
\cZ_{AHM} \propto \cZ_{BKT} =
\sum_{\stackrel{\dual j \in \Z(\dual c_3)}{\delta \dual j = 0}}
\sum_{\stackrel{\dual \sigma_j \in \Z(\dual c_2)}{\delta \dual \sigma_j =q \dual j}}
\exp\Bigl\{\!\!\!\! & - & \!\!\!\! 4 \pi^2 \beta \,
\left(j, {(\Delta + m^2)}^{-1} j\right)
\label{ZAHM2}\\
\!\!\!\! & - & \!\!\!\! 4 \pi^2 \gamma \,
\left(\sigma_j, {(\Delta + m^2)}^{-1} \sigma_j\right)
\Bigr\}\,, \nonumber
\eeqn
where we have introduced the new dual surface variable
$\dual \sigma_j = \dual \sigma + q \dual m[j]$ which is spanned
$q$--times on the monopole current $j$:
$\delta \dual \sigma_j =q \dual j$. The flux of the unit charged
magnetic monopole can be taken out by $q$ strings carrying the unit
flux. The mass of the gauge boson $\theta$ is
$m = q \sqrt{\gamma \slash \beta}$.

The BKT--representation \eq{ZAHM2} of the AHM partition function
\eq{ZAHM} can be also transformed to the dual representation using
simple Gaussian integrations. We use two dual compact fields $\dual B$
(vector field) and $\dual \xi$ (scalar field) in order to represent the
closeness properties of the currents $\dual \sigma_j$ and $\dual
j$, respectively. We also introduce two dual non--compact fields, $\dual F$
(vector field) and $\dual G$ (rank-2 tensor field) in order to get
a linear dependence, correspondingly, on the currents $\dual \sigma_j$ and
$\dual j$ under the exponential function:
\beqn
& & \cZ_{BKT} = {\mathrm {const.}} \,
\int^\infty_{-\infty} \dD \dual F \, \int^\infty_{-\infty} \dD \dual G
\int^\pi_{-\pi} \dD \dual B \, \int^\pi_{-\pi} \dD \dual \xi
\nonumber \\
& & \sum_{\dual j \in \Z(\dual c_3)} \sum_{\dual \sigma_j \in \Z(\dual c_2)}
\!\!\!\! \exp\Bigl\{ - \dual \beta  \, (\dual G, (\Delta + m^2) \dual G)
            - \dual \gamma \, (\dual F, (\Delta + m^2) \dual F)
\nonumber \\
& & \ \qquad + i (\dual F, \dual \sigma_j) + i (\dual G, \dual j)
+ i (\dual B, \delta \dual \sigma_j - q \dual j)
- i (\dual \xi, \delta \dual j) \Bigr\}\,,
\eeqn
where
\beqn
\dual \beta = \frac{1}{16 \pi^2 \gamma}\,,\quad
\dual \gamma = \frac{1}{16 \pi^2 \beta}\,.
\eeqn
Note that in this representation the integer variables $\dual \sigma_j$
and $\dual j$ are no more restricted by the closeness relations.
Therefore we can use the Poisson summation formula with respect
to these variables and integrate out the fields $\dual F$ and $\dual
G$. Finally, we obtain the dual {\it field} representation of the partition
function \eq{ZAHM}:
\beqn
\cZ_{BKT} & \propto & \cZ_{dual\, field} =
\int^\pi_{-\pi} \dD \dual B \, \int^\pi_{-\pi} \dD \dual \xi
\sum_{\dual u \in \Z(\dual c_3)} \sum_{\dual v \in \Z(\dual c_2)}
\nonumber \\
& & \exp\Bigl\{ - \dual \beta  \, \left(\dd \dual B + 2 \pi \dual u,
(\Delta + m^2) \, (\dd \dual B + 2 \pi \dual u)\right)
\label{ZAHMdual} \\
& - & \dual \gamma \, \left(\dd \dual \xi + q \dual B + 2 \pi \dual v,
(\Delta + m^2) \, (\dd \dual \xi + q \dual B + 2 \pi \dual v)\right)
\Bigr\}\,, \nonumber
\eeqn
where $\dual u$  and $\dual v$ are the integer valued forms
defined on the plaquettes and links of the dual lattice,
respectively. Clearly, this is the dual Abelian Higgs model with
the modified action. The gauge field $\dual B$ is compact and the
radial variable of the Higgs field is frozen. The model is in
the London limit and the dynamical scalar variable is
the phase of the Higgs field $\dual \xi$.

Thus in the presence of the dynamical matter the dual gauge field
$\dual B$ becomes compact\footnote{Another way to establish this fact is
to realize that the pure compact gauge model is dual to the
non--compact $U(1)$ with matter fields (referred above as the
(dual) Abelian Higgs model). Reading this relation backwards one
can conclude that the presence of the matter field leads to the
compactification of the dual gauge field~$\dual B$.}.
The compactness of the dual gauge field implies that it is
transforming under the gauge transformations of the following
form:
\beq
\dual B \rightarrow \dual B + \dd \dual \alpha + 2 \pi \dual k\,,
\label{cgauge}
\eeq
where the integer valued field $k$ is chosen in such a way that
$\dual B \in (-\pi,\pi]$.

One can easily check that the operator \eq{Phi} is not invariant under
these gauge transformations:
\beq
\Phi^{\mathrm{mon}}_x (H) \to \Phi^{\mathrm{mon}}_x (H)
\, e^{2 \pi i (\dual k, \dual H_x)}\,.
\label{change}
\eeq
The invariance of the operator \eq{Phi} under the gauge
transformations \eq{cgauge} can be achieved if and only if
the function $\dual H_x$ is an integer--valued form.

Thus, if we take into account the
Maxwell equation $\delta \dual H_x = \dual \delta_x$, we find
that $\dual H_x$ should
be a string attached to the monopole ("Mandelstam string"): $\dual H_x \to
\dual j_x$, $\dual j_x \in \Z$, $\delta \dual j_x = \dual \delta_x$. The string
must belong to the three--dimensional time--slice. However, one can
show~\cite{FrMa99} that for a fixed string position the operator $\Phi$
creates a state with an infinite energy. This difficulty may be
bypassed~\cite{FrMa99}
by summation over all possible positions of the Mandelstam strings with
a measure $\mu(\dual j)$:
\beqn
\Phi^{\mathrm{mon,new}}_x = \phi_x \,
\sum_{\stackrel{j_x \in \Z}{\delta \dual j_x = \dual \delta_x}} \mu(\dual j_x)
\, e^{i (\dual B, \dual j_x)}\,.  \label{Phi:new}
\eeqn
If Higgs field $\phi$ is
$q$--charged ($q \in \Z$), the summation in eq.\eq{Phi:new} should be
taken over $q$ different strings with the unit flux.

An example of a ``reasonable'' measure $\mu(j_x)$ is~\cite{FrMa99}:
\beqn
\mu(\dual j_x) = \exp\Bigl\{ - \frac{1}{2 \kappa}
{||\dual j_x||}^2\Bigr\}\,.  \label{Good:mu}
\eeqn
This measure corresponds to the dual formulation of
the $3D$ XY--model with the Villain action:
\beq
S(\chi, r) = {\kappa \over 2} || \dd \chi -2\pi B + 2 \pi r||^2
\,.
\label{xy}
\eeq
Due to the compactness of the spin variables $\chi$ the model \eq{xy} possesses
vortex defects which enter the XY--partition function
with measure \eq{Good:mu}.

We thus defined ``old'' \eq{Phi} and ``new'' \eq{Phi:new} monopole
creation operators.

\section{Numerical results}

Below we present results of the numerical simulation of the new
monopole creation operator. We investigate it in the simplest model
which contains both the monopoles and the electrically charged fields: the
Abelian Higgs model with compact gauge field and with the potential
on the Higgs field corresponding to the London limit. The
partition function for this model is given in eq.\eq{ZAHM}. The
model has a nontrivial phase structure and we study both the phase
where monopoles are condensed and the phase where monopoles are not
condensed.

First we substitute the monopole creation operator
(\ref{Phi:new},\ref{Good:mu}) into the dual representation of the
compact AHM~\eq{ZAHMdual}. Then we perform the transformations back to
the original representation:
\beqn
\langle\Phi^{\mathrm{mon,new}} \rangle ={1 \over Z}
\sum_{\stackrel{\dual j \in \Z(\dual c_3)}{\delta \dual j = \dual \delta_x}}
\int_{- \pi}^{ \pi} D \theta \exp \{ -{1 \over 2 \kappa} ||\dd \dual j||^2 -
\nonumber \\
- \beta \cos \left(\dd \theta + {2 \pi \over q}  \tilde j\right)
- \gamma \cos (q \, \theta)\}\,,
\eeqn
where we used the Wilson form of the action which is more suitable for
the numerical simulations. The current $\tilde j
\equiv \dual{}^{(3)} \dual{}^{(4)} j$ means that duality operation was
first applied in the 3D time slice and then in the full 4D space.
We have fixed the unitary gauge therefore the Higgs field was eaten up by the
corresponding gauge transformation.

The value of the monopole order parameter, $\langle \phi \rangle$,
corresponds to the minimum of the
(effective constraint) potential on the monopole field.
This potential can be estimated as follows:
\beqn
V_{\mathrm{eff}}(\Phi) = - \ln \Bigl(\langle \delta(\Phi
- \Phi^{\mathrm{mon,new}}) \rangle\Bigr)\,.
\label{eff:potential}
\eeqn

We simulated the 4D Abelian Higgs model on the $4^4,6^4,8^4$
lattices, for $\gamma = 0.3$. The larger the charge of the Higgs
field, $q$, the easier the numerical calculation of
$V_{\mathrm{eff}}(\Phi)$ is. We performed our calculations for
$q=7$. For each configuration of 4D fields we simulated 3D model
to get the ensemble of the Mandelstam strings with the weight
$\mu(j_x)$. We generated 60 statistically independent 4D field
configurations, and for each of these configurations we generated
40 configurations of 3D Mandelstam strings. We imposed the
anti-periodic boundary conditions in the $3D$ space (the single
monopole charge can not exist in the finite volume with periodic
boundary condition).

As we have stated above the weight function \eq{Good:mu}
corresponds to the 3D XY--model with the Villain action. This
model has the phase transition at $\kappa_c(B=0) \approx
0.32$~\cite{Kog77}. Our numerical observation has shown that in
presence of the external field $B$, the critical coupling constant
gets shifted: $\kappa_c^B \approx 0.42$.

\begin{figure}[!htb]
\begin{center}
\begin{tabular}{cc}
\epsfxsize=5.2cm \epsffile{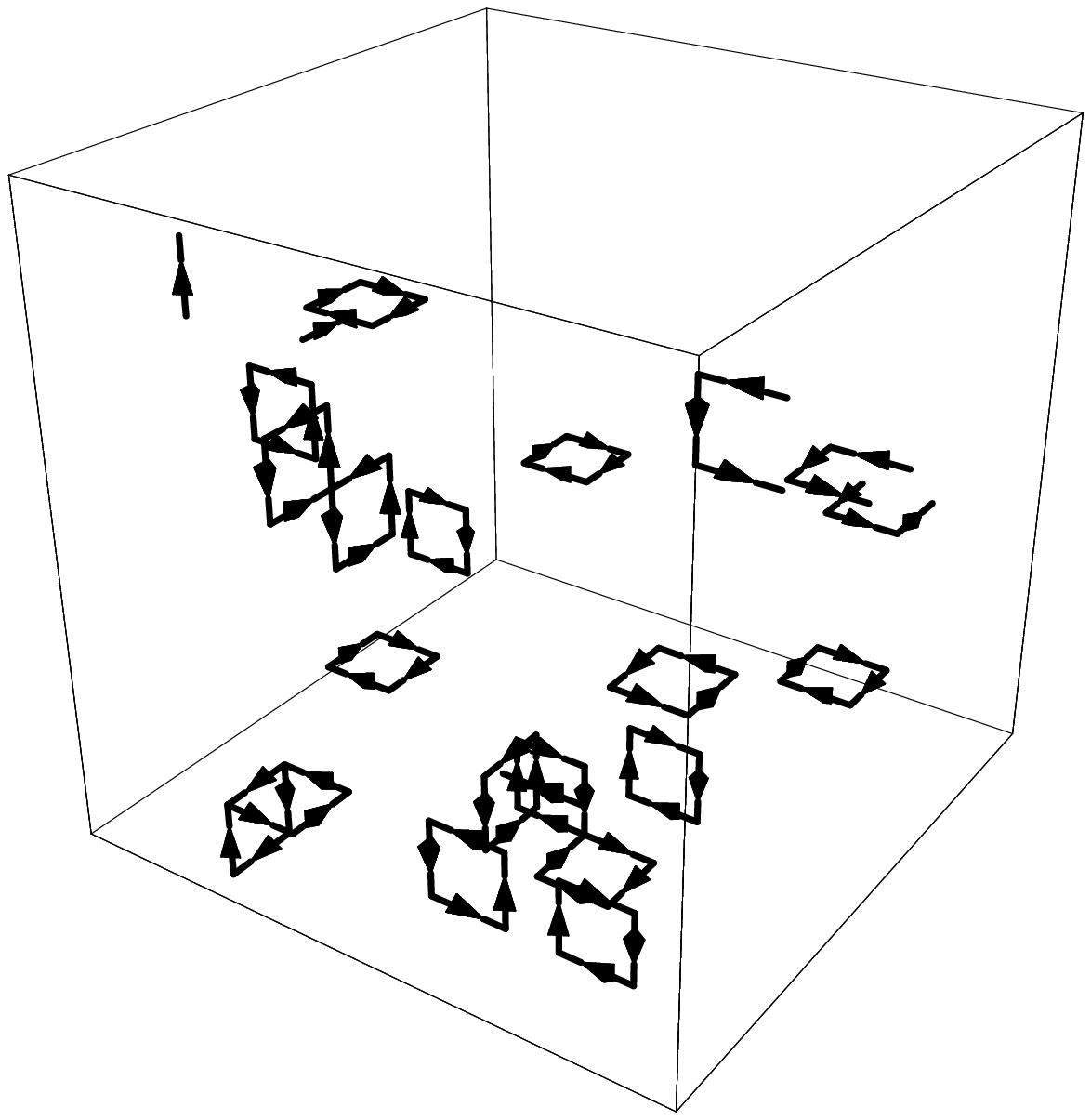} &
\epsfxsize=5.2cm \epsffile{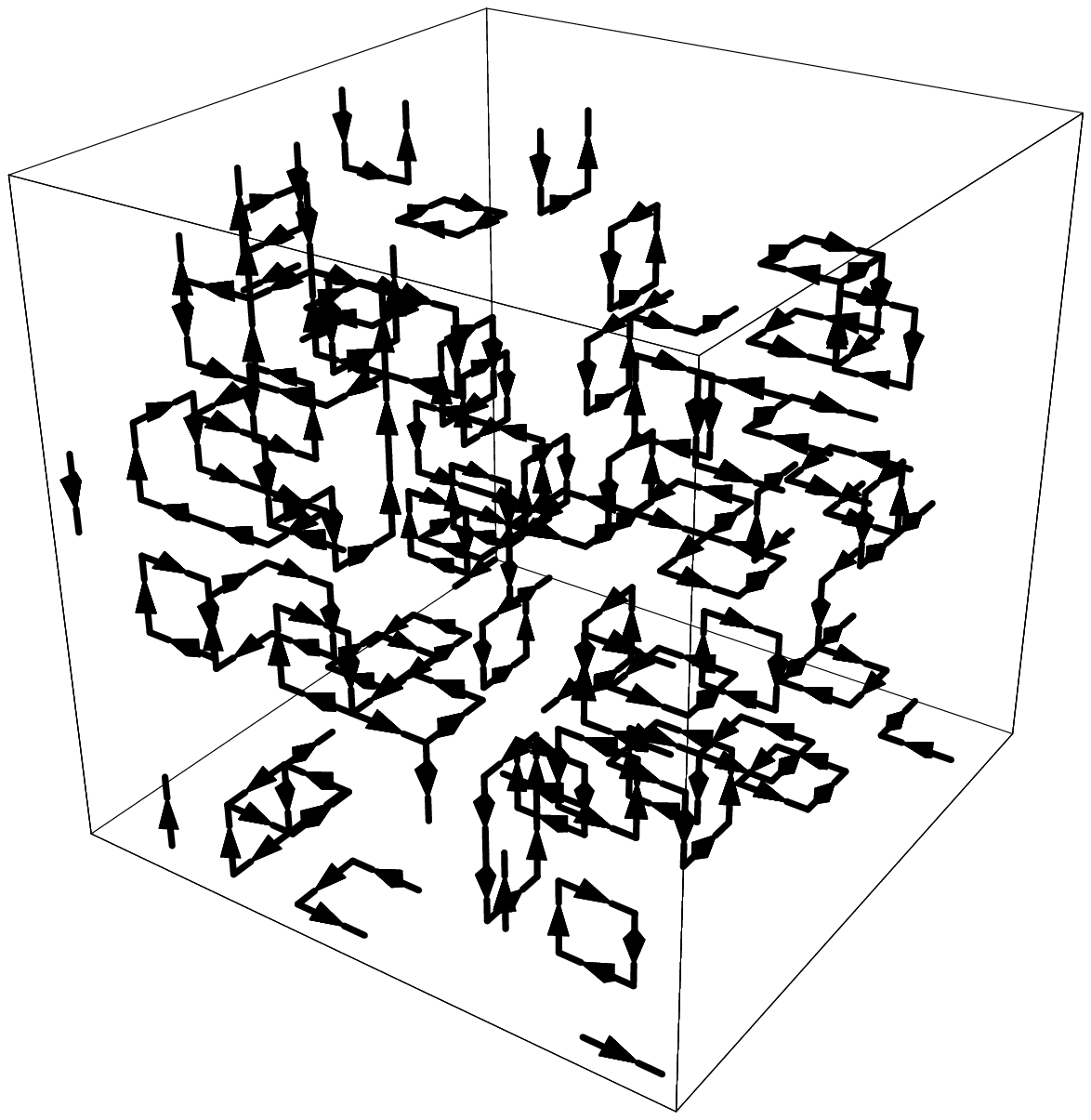} \\
(a) &
(b)
\end{tabular}
\end{center}
\caption{Charged loops appearing in auxiliary theory for (a)
non--condensed strings ($\kappa=0.3$) and (b)
condensed strings ($\kappa=0.5$).}
\label{fig:loops}
\end{figure}

Two configurations of the Man\-del\-stam strings which correspond
to condensed (large $\kappa$) and non--condensed (small $\kappa$)
phases of these strings are shown in Figures~\ref{fig:loops}.

We can expect that the operator~\eq{Phi:new} plays the role of the order
parameter in the $\kappa>\kappa_c$ phase, where the Mandelstam strings
are condensed ($\kappa>\kappa_c$).
\begin{figure}[!htb]
\begin{center}
\begin{tabular}{cc}
\epsfxsize=6.cm \epsffile{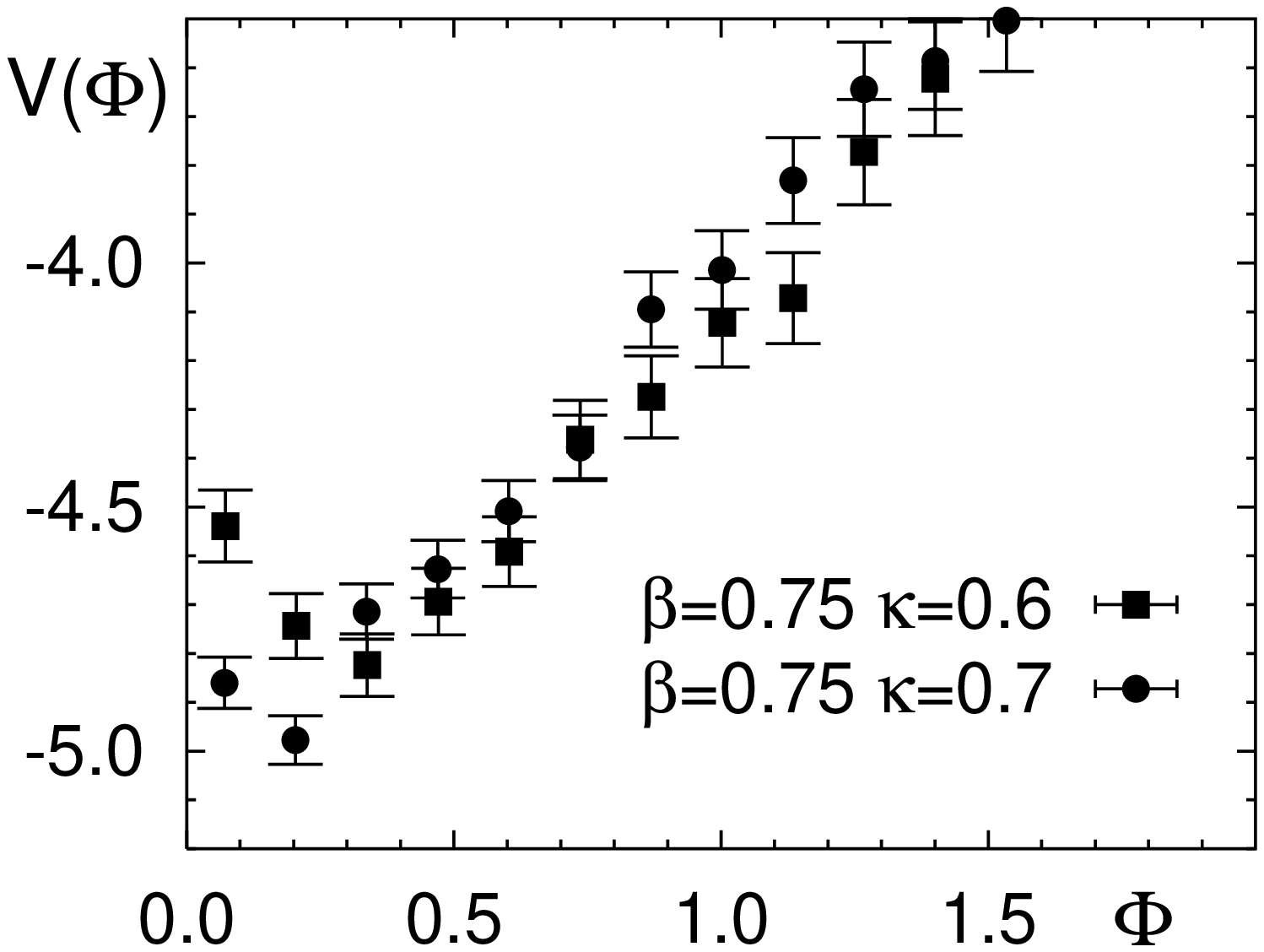} &
\epsfxsize=6.cm \epsffile{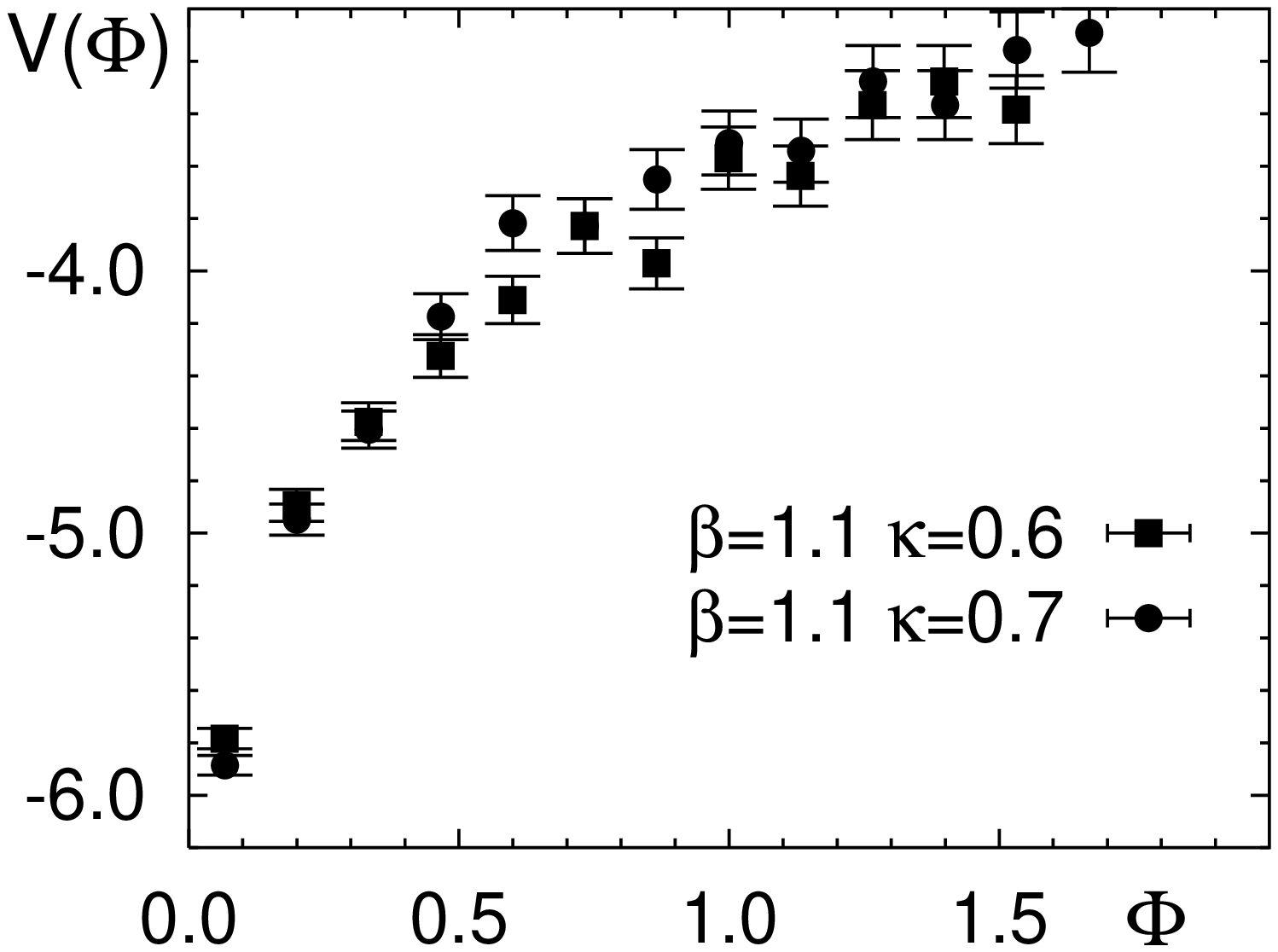} \\
(a) &
(b)
\end{tabular}
\end{center}
\caption{The effective monopole potential~\eq{eff:potential}
in (a) confinement and (b) deconfinement phases.}
\label{fig:potentials}
\end{figure}
In Figures~\ref{fig:potentials} we
present the effective potential \eq{eff:potential} in the confinement
($\beta=0.85$) and deconfinement ($\beta=1.05$) phases. The potential is
shown for two values of the $3D$ coupling constants $\kappa > \kappa_c$
corresponding to high densities of the Mandelstam strings. In the
confinement phase, Figure~\ref{fig:potentials}(a), the potential
$V(\Phi)$ has a Higgs form signaling the monopole
condensation. According to our numerical observations this statement
does not depend on the lattice volume. In the deconfinement phase,
Figure~\ref{fig:potentials}(b), the potential has minimum at $\Phi = 0$
which indicates the absence of the monopole condensate.

\begin{figure}[!htb]
\vspace{2mm}
\begin{center}
\epsfxsize=8.0cm \epsffile{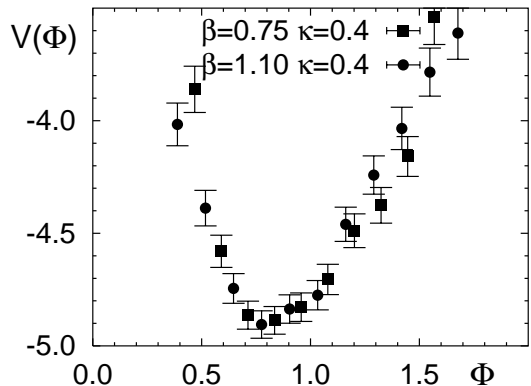}
\end{center}
\caption{The effective monopole potential~\eq{eff:potential}
in the low--$\kappa$ region of the $3D$ model.}
\label{fig:smallk}
\end{figure}
For small values of the $3D$ coupling constant $\kappa$ (in the phase
where Mandelstam strings $j_x$ are not condensed), we found
(Figure~\ref{fig:smallk}) that the potential $V(\Phi)$ has the same behaviour
for the both phases of $4D$ model. Thus the operator \eq{Phi:new} serves as
the order parameter for the deconfinement phase transition, if
Mandelstam strings are condensed, i.e. $\kappa$ should be larger
than $\kappa_c (B)$.

Summarizing, the new operator suggested in \cite{FrMa99} can be used as
a test of the monopole condensation in the theories with electrically
charged matter fields. Our calculations indicate that the operator
should be defined in the phase where the Mandelstam strings are
condensed. The value of the monopole field which corresponds to the
minimum of the effective potential, is zero in deconfinement phase and
non zero in the confinement phase.

\newpage

\section*{Acknowledgments}

The authors are grateful to F.V. Gubarev for useful
discussions. V.A.B. and M.I.P were partially supported by grants RFBR
02-02-17308, RFBR 01-02-17456, INTAS 00-00111 and CRDF award
RP1-2364-MO-02.  M.I.Ch. is supported by JSPS Fellowship P01023.


\begin{thebibliography}{99}

\bibitem{DualSuperconductor}
't~Hooft, G. (1975) in {\it High Energy Physics}, ed. A. Zichichi,
EPS International Conference, Palermo;
Mandelstam, S. (1976)
``Vortices And Quark Confinement In Nonabelian Gauge Theories'',
{\it Phys.\ Rept.}  {\bf 23}, 245.

\bibitem{AbelianProjections}
't Hooft, G.  (1981)
``Topology Of The Gauge Condition And New Confinement Phases In Nonabelian Gauge
Theories'', {\it Nucl.\ Phys.} {\bf B190}, 455;\\

\bibitem{FrMa87}
Fr\"ohlich, J. and Marchetti, P.~A.,
``Soliton Quantization In Lattice Field Theories'',
Commun.\ Math.\ Phys.\  {\bf 112} (1987) 343.

\bibitem{PoWi} Polley, L. and Wiese, U.~J. (1991)
``Monopole Condensate And Monopole Mass In U(1) Lattice Gauge
Theory'', {\it Nucl.\ Phys.\ B} {\bf 356}, 629;\\
Polikarpov, M.~I., Polley, L. and Wiese, U.~J. (1991)
``The Monopole Constraint Effective Potential In U(1) Lattice
Gauge Theory'', {\it Phys.~Lett.} {\bf B} {\bf 253} (1991) 212.

\bibitem{ChPoVe}
Chernodub, M.~N., Polikarpov, M.~I. and Ve\-se\-lov, A.~I. (1996)
``Monopole Order Parameter In SU(2) Lattice Gauge Theory'',
JETP Lett.\  {\bf 63}, 411;
%
``Effective constraint potential for Abelian monopole in SU(2) lattice  gauge
theory'', {\it Phys.~Lett.} {\bf B399}, 267;
%
``Monopole order parameter in SU(2) lattice gauge theory''
{\it Nucl. Phys. Proc. Suppl.}  {\bf 49}, 307;
%
``Effective monopole potential for SU(2) lattice gluodynamics in spatial
maximal Abelian gauge'', {\it JETP Lett.}  {\bf 69}, 174.

\bibitem{DiGi} Di Giacomo, A. and Paffuti, G. (1997)
``A disorder parameter for dual superconductivity in gauge
theories'', {\it Phys.~Rev. D} {\bf 56} 6816;\\
%
Nakamura~N.,~et~al. ``Disorder parameter of confinement'' (1997),
{\it Nucl. Phys. Proc. Suppl.} {\bf 53} 512.

\bibitem{FrMa99}
Fr\"ohlich, J. and Marchetti, P.~A. (1999)
``Gauge-invariant charged, monopole and dyon fields in gauge
theories'', {\it Nucl.\ Phys. B} {\bf 551}, 770;\\
%
Fr\"ohlich, J. and Marchetti, P.~A. (2001)
``An order parameter reconciling Abelian and center
dominance in SU(2) Yang-Mills theory'',
{\it Phys. Rev. D} {\bf 64}, 014505

\bibitem{Dirac} Dirac, P.A.M. (1955)
``Gauge invariant formulation of quantum electrodynamics'',
{\it Can.~J.~Phys}, {\bf 33}, 650.

\bibitem{Review}
Chernodub, M.~N. and Polikarpov, M.~I. (1997)
``Abelian projections and monopoles'',
in "Confinement, duality, and nonperturbative aspects of QCD",
Ed. by P. van Baal, Plenum Press, p. 387, hep-th/9710205.

\bibitem{BKT}
Berezinsky, V.~L. (1971)
"Destruction Of Long Range Order In One-Dimensional And
Two-Dimensional Systems Having A Continuous Symmetry Group.
1. Classical Systems",
Sov.\ Phys.\ JETP {\bf 32}, 493.
%
Kosterlitz, J.M. and Thouless, D.J. (1973), {\it J. Phys.} {\bf C6},
1181; J.M. Kosterlitz, J. Phys. {\bf C 7} (1974) 1046.

\bibitem{Kog77}
Banks, T., Myerson, R. and Kogut, J., (1977)
``Phase Transitions In Abelian Lattice Gauge Theories'',
{\it Nucl.~Phys. B} {\bf 129}, 493.

\end{thebibliography}
\end{document}